\documentclass[11pt, a4paper, center, fleqn, oneside]{report} 
\usepackage[italian]{babel} 
\usepackage[dvips]{graphicx} 
\usepackage{indentfirst} 
\usepackage{longtable} 
\usepackage{lscape} 
\begin{document} 

\begin{center} 
\section*{GOLD Mine: A new Galaxy Database on the WEB}

\emph{Gavazzi G., Boselli A., Donati A., Franzetti P., Scodeggio M.}

\end{center}
The galaxy database "GOLDmine" (http://goldmine.mib.infn.it/) has been significantly updated (Sept/1/2003)
\noindent
(see "Introducing GOLD Mine: A new Galaxy Database on the WEB" by Gavazzi et al. 2003, Astronomy \& Astrophysics, 400, 451).\\
\noindent
The new features include:\\
a) Sample extension:
\begin{enumerate}
\item the GOLDmine sample has been extended from the original Virgo cluster + Coma supercluster regions to
include the clusters: A262, Cancer, A2147, A2151, A2197, A2199. 382 galaxies from the GCGC (with $m_p<15.7$)
have been added in these regions.
\end{enumerate}
b) New query keys:
\begin{enumerate}
\item query by near position (and near name).
\item query by available images.
\end{enumerate}
c) Routinary image update:
\begin{enumerate}
\item  59 (B). 72 (V) and 70 ($\rm H_{\alpha}$) new frames from observations carried on by the GOLDmine team in spring 2003.
\item 157 new optical (drift-scan) spectra from  observations carried on by the GOLDmine team in 2002-2003.
\item 225 B frames of VCC galaxies taken with the INT (kindly provided by S. Sabatini).
\item 56 B frames of galaxies in A1367 taken with the CFHT (kindly provided by M. Treyer).
\item 20 (H), 32 (K) band frames of bright Virgo members (from 2MASS).
\end{enumerate}
The new numbers in GOLDmine are:
\begin{itemize}
\item galaxies:		      		3649
\item V-band frames:  	      		 706
\item B-band frames:				 858
\item $\rm H_{\alpha}$ frames (NET):     		 385
\item $\rm H_{\alpha}$ frames (OFF-band):    		 385
\item H-band frames:  	      		1241
\item K-band frames:  	      		 114
\item Spectra:		      		 323
\end{itemize}
All frames are available in FITS (and jpg) format.
\end{document}